\documentclass[seceqn,dvips]{arxstspdf}
\usepackage{graphicx}
\usepackage{flushend}
\usepackage{stfloats}

\volume{25}
\issue{4}
\pubyear{2010}
\firstpage{533}
\lastpage{544}
\doi{10.1214/10-STS348}

\makeatletter
\newcommand{\Xobs}{X_{\mathrm{obs}}}
\newcommand{\Xmis}{X_{\mathrm{mis}}}
\newcommand{\Prob}[1]{\operatorname{Pr}(#1)}
\newcommand{\cal}{\mathcal}
\newcommand{\eqref}[1]{(\ref{#1})}
\makeatother

\begin{document}
\begin{frontmatter}

\title{Parameter Expansion and Efficient Inference}
\runtitle{Parameter Expansion}

\begin{aug}
\author[a]{\fnms{Andrew} \snm{Lewandowski}\ead[label=e1]{alewand@purdue.edu}},
\author[b]{\fnms{Chuanhai} \snm{Liu}\corref{}\ead[label=e2]{chuanhai@purdue.edu}%
\ead[label=u1,url]{www.stat.purdue.edu}}
\and
\author[c]{\fnms{Scott} \snm{Vander Wiel}\ead[label=e3]{scottv@lanl.gov}%
\ead[label=u2,url]{www.stat.lanl.gov}}
\runauthor{A. Lewandowski, C. Liu and S. Vander Wiel}

\affiliation{Purdue University and Los Alamos National Laboratory}

\address[a]{Andrew Lewandowski is Ph.D. Student,
Department of Statistics, Purdue University,
150 N. University Street, West Lafayette, Indiana 47907, USA
\printead{e1}.}
\address[b]{Chuanhai Liu is Professor of Statistics,
Department of Statistics, Purdue University,
150 N. University Street, West Lafayette, Indiana 47907, USA
\printead{e2,u1}.}
\address[c]{Scott Vander Wiel is Technical Staff Member,
Statistical Sciences Group, MS F600, Los~Alamos National
Laboratory, Los Alamos, New Mexico 87545, USA
\printead{e3,u2}.}

\end{aug}

%
\begin{abstract}
This EM review article focuses on parameter expansion, a simple technique
introduced in the PX-EM algorithm to make EM converge faster while
maintaining its simplicity and stability. The primary objective concerns
the connection between parameter expansion and efficient inference.
It reviews the statistical interpretation of the PX-EM algorithm,
in terms of efficient inference via bias reduction, and further unfolds
the PX-EM mystery by looking at PX-EM from different perspectives.
In addition, it briefly discusses potential applications of parameter expansion
to statistical inference and the broader impact of statistical thinking
on understanding and developing other iterative optimization algorithms.
\end{abstract}

%

\end{frontmatter}

\section{Introduction}
\label{sec:intro}

The expectation maximization (EM) algorithm of Dempster, Laird and Rubin (\citeyear{DLR})
has proven
to be a popular computational
method for optimization.
While simple to implement and stable in its convergence, the EM
algorithm can
converge slowly.
Many variants of the original EM algorithm have also been proposed in
the last 30$+$ years
in order to overcome shortcomings that are sometimes seen in
implementations of
the original method.
Among these EM-type algorithms are the expectation-conditional
maximization (ECM) algorithm of Meng and Rubin (\citeyear{MengRubin1993}),
the expectation-conditional maximization either (ECME) algorithm of
Liu and Rubin (\citeyear{LiuRubin1994}),
the alternating ECM (AECM) algorithm of Meng and van Dyk (\citeyear{MengVanDyk1997}) %
and, more recently, the dynamic ECME (DECME) algorithm of He and Liu (\citeyear{he2009dynamic}).
This review article focuses on parameter expansion as a way of
improving the performance
of the EM algorithm through a discussion of the parameter expansion EM
(PX-EM) algorithm proposed by Liu, Rubin and Wu (\citeyear{LiuRubinWu1998}).

The EM algorithm is an iterative algorithm for maximum likelihood (ML)
estimation from incomplete
data. Let $\Xobs$ be the observed data and let
$f(\Xobs;\theta)$ denote the observed-data
model with unknown parameter $\theta$, where
$\Xobs\in{\cal X}_{\mathrm{obs}}$ and $\theta\in\Theta$.
Suppose that the observed-data model can be obtained from
a complete-data model, denoted by\break
$g(\Xobs, \Xmis;\theta)$, where
$\Xobs\in{\cal X}_{\mathrm{obs}}$,
$\Xmis\in{\cal X}_{\mathrm{mis}}$,
and $\theta\in\Theta$. That is,
\[
f(\Xobs;\theta) = \int_{{\cal X}_{\mathrm{mis}}}
g(\Xobs, \Xmis;\theta)\,d\Xmis.
\]
Given a starting point $\theta^{(0)}\in\Theta$, the EM algorithm
iterates for $t=0,1,\ldots  $ between the
expectation (E) step and maximization (M) step:
\begin{description}
\item[\textnormal{\textit{E step.}}] Compute the expected complete-data log-like\-lihood
%
\begin{eqnarray}\label{eq:Q-fun}
&&Q\bigl(\theta|\theta^{(t)}\bigr)\nonumber
\\[-8pt]
\\[-8pt]
&& \quad = \mathrm{E} \bigl(\ln g(\Xobs, \Xmis;\theta)
|\Xobs, \theta=\theta^{(t)} \bigr)
\nonumber
\end{eqnarray}
as a function of $\theta\in\Theta$; and
\item[\textnormal{\textit{M step.}}] Maximize $Q(\theta|\theta^{(t)})$ to obtain
%
\begin{equation}\label{eq:maxQ-fun}
\theta^{(t+1)} = \arg\max_{\theta\in\Theta}Q\bigl(\theta|\theta^{(t)}\bigr).
\end{equation}
\end{description}
Two EM examples are given in Section \ref{sec:em-examples}.

Roughly speaking, the E step can be viewed as creating a
complete-data problem by imputing missing values,
and the M step can be understood as
conducting a maximum likelihood-based analysis.
More exactly, for complete-data models belonging to the exponential family,
the E step imputes the complete-data sufficient statistics with their
conditional expectations given the observed data and the current estimate
$\theta^{(t)}$ of the parameter $\theta$.
This to some extent explains the simplicity of EM.
The particular choice of (\ref{eq:Q-fun}) together with Jensen's inequality
implies monotone convergence of EM.

The PX-EM algorithm is essentially an EM algorithm,
but it performs inference on a larger full model. This model is
obtained by
introducing extra parameters into the complete-data model while preserving
the observed-data sampling model.
Section \ref{sec:px-em-def} presents the structure used in PX-EM.
The theoretical results established by Liu, Rubin and Wu (\citeyear{LiuRubinWu1998})
show that PX-EM converges no slower than its parent EM.
This is somewhat surprising, as it is commonly believed that
optimization algorithms
generally converge slower as the number of dimensions increases.
To help understand the behavior of PX-EM, Liu, Rubin and Wu (\citeyear{LiuRubinWu1998})
provided a statistical interpretation
of the PX-M step in terms of
covariance adjustment. This is reviewed in Section \ref{sec:px-robit} %
in terms of bias reduction using the example of
binary regression with a Student-$t$ link (see
Mudholkar and George, \citeyear{MudholkarGeorge1978}; Albert and Chib, \citeyear{AlbertChib1993};
Liu, \citeyear{Liu2004}), which
serves as a simple robust alternative to logistic regression
and is called robit regression by Liu (\citeyear{Liu2004}).

To help further understand why PX-EM can work so well, several relevant
issues are discussed in
Section \ref{sec:unfolding}.
Section \ref{sec:how-to} provides additional motivation behind why
PX-EM can improve upon EM or ECM.
In Section \ref{sec:eda} we argue that parameter expansion
can also be used for efficient data augmentation in the E step.
The resulting EM is effectively the PX-EM algorithm.

In addition to the models discussed here,
parameter expansion has now been shown to have computational advantages in
applications such as factor analysis (Liu, Rubin and Wu, \citeyear{LiuRubinWu1998})
and the analysis of both linear (Gelman et al., \citeyear{Gelman2008}) and nonlinear
(Lavielle and Meza, \citeyear{Lavielle2007})
hierarchical models.
However, Gelman (\citeyear{Gelman2004}) shows that parameter
expansion offers more than a computational method to accelerate EM.
He points out that parameter expansion can be viewed as
part of a larger perspective on iterative simulation
(see Liu and Wu, \citeyear{LiuWu1999}; Meng and van Dyk, \citeyear{MD1999}; van Dyk and Meng,
\citeyear{vanDykMeng2001}; Liu, \citeyear{Liu2003}) and that it suggests a new family of prior
distributions in a Bayesian framework discussed by Gelman (\citeyear{Gelman2006}).
One example is the folded noncentral Student-$t$ distribution for
between-group variance parameters in hierarchical models.
This method exploits a parameter expansion technique commonly used in
hierarchical
models, and Gelman (\citeyear{Gelman2006}) shows that it can be more robust than
the more common inverse-gamma prior.
Inspired by Gelman (\citeyear{Gelman2004}), we briefly discuss other potential
applications of parameter expansion to statistical inference in
Section \ref{sec:discussion}.

\section{Two EM examples}
\label{sec:em-examples}
\subsection{The Running Example: A Simple Poisson--Binomial
Mixed-Effects Model}
\label{sec:ae}

Consider the complete-data model for the observed data $\Xobs= X$ and
the missing data $\Xmis= Z$:
\[
Z|\lambda\sim\operatorname{Poisson}(\lambda)
\]
and
\[
X|(Z, \lambda) \sim\operatorname{Binomial}(Z, \pi),
\]
where $\pi\in(0,1)$ is known and $\lambda>0$ is the unknown parameter
to be estimated.

The observed-data model $f(X;\lambda)$ is obtained from the joint
sampling model
of $(X,Z)$:
%
\begin{eqnarray} \label{eq:ae-likelihood}
&&g(X,Z;\lambda)\nonumber
\\[-8pt]
\\[-8pt]
&& \quad = \frac{\lambda^Z}{Z!}e^{-\lambda}\frac{Z!}{X!(Z-X)!}
\pi^X(1-\pi)^{Z-X},
\nonumber
\end{eqnarray}
where $X=0, 1,\ldots  , Z$, $Z=0, 1, \ldots ,$ and $\lambda\geq0$.
That is, $f(X;\lambda)$ is derived from $g(X,Z;\lambda)$
by integrating out the missing data $Z$ as follows:
\begin{eqnarray*}
f(X;\lambda) & = &\sum_{z=X}^\infty\frac{\lambda^z}{z!}e^{-\lambda
}\frac{z!}{X!(z-X)!}
\pi^X(1-\pi)^{z-X}\\
&=& \frac{\lambda^{X}\pi^X}{X!}e^{-\lambda} \sum_{z=X}^\infty
\frac{\lambda^{z-X}}{(z-X)!} (1-\pi)^{z-X}\\
&\stackrel{k=z-X}{=}& \frac{(\lambda\pi)^{X}}{X!}e^{-\lambda} \sum
_{k=0}^\infty
\frac{[\lambda(1-\pi)]^{k}}{k!}\\
&=& \frac{(\lambda\pi)^{X}}{X!}e^{-\lambda} e^{\lambda(1-\pi)}\\
&=& \frac{(\lambda\pi)^{X}}{X!}e^{-\lambda\pi}.
\end{eqnarray*}
Alternatively, one can get the result from
the well-known fact related to
the infinite divisibility of the Poisson distribution; namely,
if $X_1=X$ and $X_2=Z-X$ are independent
Poisson random variables with rate $\lambda_1=\lambda\pi$ and
$\lambda_2=\lambda(1-\pi)$, then
$X_1+X_2\sim\operatorname{Poisson}(\lambda_1+\lambda_2)$ and
conditional on $X_1+X_2$, $X_1\sim\operatorname{Binomial}(X_1+X_2, \lambda
_1/(\lambda_1+\lambda_2))$.

It follows that the observed-data model is
$X|\lambda\sim\operatorname{Poisson}(\pi\lambda)$.
Thus, the ML estimate of $\lambda$ has a closed-form solution,
$\hat\lambda= X/\pi.$
This artificial example serves two purposes.
First, it is easy
to illustrate the general EM derivation.
Second, we use this example in Section \ref{sec:ae-px-em}
to show an extreme case in which PX-EM can converge
dramatically faster than its parent EM; PX-EM converges in one-step,
whereas EM converges painfully slowly.

The complete-data likelihood is given by
the joint sampling model of $(X,Z)$ found in equation \eqref{eq:ae-likelihood}.
It follows that the complete-data model belongs to the exponential
family with sufficient statistic $Z$ for $\lambda$.
The complete-data ML estimate of $\lambda$ is given by
%
\begin{equation}\label{eq:lh-mle-toy}
\hat\lambda_{\mathrm{com}} = Z.
\end{equation}
To derive the E step of EM, the conditional distribution of
the missing data $Z$ given both the observed data and
the current estimate of the parameter $\lambda$ is
used. It is determined as follows:
\begin{eqnarray*}
h(Z|X,\lambda) &=& \frac{g(X,Z;\lambda)}{\sum_{z=X}^\infty
g(X,z;\lambda)}\\
&=& \frac{  [\lambda(1-\pi)]^{Z-X}/(Z-X)!}
{\sum_{z=X}^\infty ([\lambda(1-\pi)]^{z-X}/(z-X)!) }\\
&=& \frac{[\lambda(1-\pi)]^{Z-X}}{(Z-X)!} e^{\lambda(1-\pi)}.
\end{eqnarray*}
Thus,
$
Z|\{X,\lambda\} \sim X+ \operatorname{Poisson}(\lambda(1-\pi)).
$
This yields
\[
E (Z|X,\lambda ) = X + \lambda(1-\pi).
\]
Thus, the EM algorithm follows from the discussion of Dempster, Laird and Rubin (\citeyear{DLR})
on exponential complete-data models. Specifically,
given the updated estimate $\lambda^{(t)}$ at the $t$th iteration,
EM follows these two steps:
\begin{description}
\item[\textnormal{\textit {E step.}}] Compute $\hat{Z} = \mathrm{E}(Z|X, \lambda=\lambda^{(t)})
= X + \lambda^{(t)}\times\break (1-\pi)$.
\item[\textnormal{\textit {M step.}}] Replace $Z$ in (\ref{eq:lh-mle-toy})
with $\hat{Z}$ to obtain $\lambda^{(t+1)} = \hat{Z}$.
\end{description}

It is clear that the EM sequence $\{\lambda^{(t)}\dvtx  t=0,1, \ldots \}$
is given by
%
\begin{equation}\label{eq:ac-em-seq}
 \quad \lambda^{(t+1)} = X + \lambda^{(t)}(1-\pi)\quad(t=0,1,\ldots )
\end{equation}
converging to the ML estimate
\[
\hat\lambda= X/\pi.
\]
Rewrite (\ref{eq:ac-em-seq}) as
\[
\lambda^{(t+1)} - \hat\lambda= (1-\pi) \bigl(\lambda^{(t)} - \hat
{\lambda} \bigr)
\]
%
to produce a closed-form expression for the convergence rate of EM:
\[
\frac{|\lambda^{(t+1)}-\hat\lambda|}{|\lambda^{(t)}-\hat\lambda
|} = 1- \pi.
\]
This indicates that EM can be very slow when $\pi\approx0$.

\subsection{ML Estimation of Robit Regression via EM}
\label{sec:em}

Consider the observed data consisting of $n$ observations
$\Xobs=\{(x_i, y_i)\dvtx  i=1,\ldots ,n\}$ with a $p$-dimensional covariate
vector $x_i$ and binary response $y_i$ that takes on values of 0 and 1.
The binary regression model with Student-$t$ link
assumes that, given the covariates,
the binary responses $y_i$'s are independent
with the marginal probability distributions specified by
%
\begin{eqnarray}\label{eq:robit}
 \qquad \Prob{y_i=1|x_i, \beta} &=&
1-\Prob{y_i=0|x_i, \beta}\nonumber
\\[-8pt]
\\[-8pt]
 \qquad  &=& F_\nu(x_i'\beta)\quad(i=1,\ldots ,n),
\nonumber
\end{eqnarray}
where $F_\nu(\cdot)$ denotes the c.d.f. of the Student-$t$ distribution with
center zero, unit scale and $\nu$ degrees of freedom.
With $\nu\approx7$, this model provides a robust approximation to
the popular logistic regression model for binary data analysis.
Here we consider the case with known $\nu$.

The observed-data likelihood
\begin{eqnarray*}
&&f(\Xobs; \beta)\\
&& \quad  =
\prod_{i=1}^n [F_\nu(x_i'\beta)]^{y_{i}}[1-F_\nu(x_i'\beta)]^{1-y_i}
 \quad  (\beta\in{\mathbb R}^p)
\end{eqnarray*}
involves the product of the c.d.f. of the Student-$t$ distribution
$F_\nu(\cdot)$ evaluated at $x_i'\beta$ for $i=1,\ldots ,n$.
The MLE of $\beta$ does not appear to have a closed-form solution.
Here we consider the EM algorithm for finding the MLE of $\beta$.

A complete-data model for implementing EM to find the ML estimate of
$\beta$ is specified by introducing the missing data consisting of
independent latent variables
$(\tau_i, z_i)$ for each $i=1,\ldots ,n$ with
%
\begin{equation}\label{eq:tau}
\tau_i|\beta\sim\operatorname{Gamma}(\nu/2,\nu/2)
\end{equation}
and
%
\begin{equation}\label{eq:z}
z_i|(\tau_i,\beta) \sim\mathrm{N}(x_i'\beta,1/\tau_i).
\end{equation}
Let
%
\begin{equation}\label{eq:y}
y_i=
\cases{\displaystyle
1, & if $z_i>0$, \cr\displaystyle
0, & if $z_i\leq0$
}\hspace*{6pt}
 (i=1,\ldots ,n).
\end{equation}
Then the marginal distribution of $y_i$ is preserved and is given
by (\ref{eq:robit}).
The complete-data model belongs to the exponential family and has the
following sufficient statistics for $\beta$:
%
\begin{equation}\label{eq:ss}
 \qquad S_{\tau xx'} = \sum_{i=1}^n\tau_ix_ix_i'
 \quad \mbox{and} \quad
S_{\tau xz} = \sum_{i=1}^n\tau_ix_iz_i'.
\end{equation}
The complete-data ML estimate of $\beta$ is given by
%
\begin{equation}\label{eq:MLE_com}
\hat{\beta}_{{\mathrm{com}}} = S_{\tau xx'}^{-1}S_{\tau xz},
\end{equation}
leading to the following EM algorithm.

Starting with $\beta^{(0)}$, say, $\beta^{(0)}=(0,\ldots ,0)$,
EM iterates for $t=0,1,\ldots $ with
iteration $t+1$ consisting of the following
E and M steps:
\begin{description}
\item[\textnormal{\textit {E step.}}] Compute $\hat{S}_{\tau xx'}=
\mathrm{E} (S_{\tau xx'}|\beta= \beta^{(t)},\Xobs )$
and $\hat{S}_{\tau xz}=
\mathrm{E} (S_{\tau xz}|\beta= \beta^{(t)},\Xobs )$.
\item[\textnormal{\textit {M step.}}] Update the estimate of\vspace*{1pt} $\beta$ to obtain
$\beta^{(t+1)} = \hat{S}_{\tau xx'}^{-1}\hat{S}_{\tau xz}$.
\end{description}
Let $f_\nu(\cdot)$ denote the p.d.f. of $F_\nu(\cdot)$.
The E step can be coded by using the following results derived in
Liu (\citeyear{Liu2004}):
%
\begin{eqnarray}
    \hspace*{22pt}\hat{\tau}_i &=& \mathrm{E} \bigl(\tau_i|\beta= \beta^{(t)},\Xobs
 \bigr)\nonumber
 \\[-8pt]
 \\[-8pt]
  \hspace*{22pt} &=& \frac{y_i-(2y_i-1)F_{\nu+2}(-(1+2/\nu)^{1/2}x_i'\beta^{(t)})}
{y_i-(2y_i-1)F_{\nu}(-x_i'\beta^{(t)})},
\nonumber
\end{eqnarray}\vspace*{-10pt}
\begin{equation}
   \hat{\tau_iz_i} = \mathrm{E} \bigl(\tau_iz_i|\beta= \beta
^{(t)},\Xobs \bigr)  =  \hat{\tau}_i\hat{z_i},
\end{equation}
where
\begin{eqnarray*}
\hat{z}_i &\equiv& x_i'\beta^{(t)}\\
&&{} +
\frac{(2y_i-1)f_\nu(x_i'\beta^{(t)})}
{y_i-(2y_i-1)F_{\nu+2}(-(1+2/\nu)^{1/2}x_i'\beta^{(t)})}
\end{eqnarray*}
for $i=1,\ldots ,n$.

However, the EM algorithm can also converge slow\-ly in this example.
This is discussed in Section \ref{sec:px-robit},
where it is shown that PX-EM can greatly improve the convergence rate.

\section{The PX-EM Algorithm}
\label{sec:px-em}

\subsection{The Algorithm}
\label{sec:px-em-def}

Suppose that the EM complete-data model can be embedded in a larger
model
$g_*(\Xobs, \Xmis;\theta_*, \alpha)$
with the expanded parameter
$(\theta_*, \alpha)\in\Theta\times{\cal A}$.
Assume that the observed-data model is preserved in the
sense that, for every $(\theta_*, \alpha)\in\Theta\times{\cal A}$,
%
\begin{equation}\label{eq:px-preservation}
f(\Xobs;\theta)
= f_*(\Xobs;\theta_*, \alpha)
\end{equation}
holds for some $\theta\in\Theta$, where
$
f_*(\Xobs;\theta_*, \alpha) =\break
\int_{{\cal X}_{\mathrm{mis}}}
g_*(\Xobs, \Xmis;\theta_*, \alpha)\,d\Xmis.
$
The\vspace*{1pt} condition (\ref{eq:px-preservation}) defines a mapping
$\theta= R(\theta_*, \alpha)$, called
the reduction function, from the expanded parameter space
$\Theta\times{\cal A}$ to the original parameter space $\Theta$.
For convenience, assume that the expanded parameters
are represented in such a way that
the original complete-data and observed-data models are recovered by fixing
$\alpha$ at $\alpha_0$. Formally, assume that
there exists a \textit{null} value of $\alpha$, denoted by $\alpha_0$,
such that $\theta= R(\theta, \alpha_0)$ for every
$\theta\in\Theta$.
When applied to the parameter-expanded complete-data model
$g_*(\Xobs, \Xmis;\theta_*, \alpha)$, the EM algorithm,
called the PX-EM algorithm, creates a sequence
$\{(\theta_*^{(t)}, \alpha^{(t)})\}$ in $\Theta\times{\cal A}$.
In the original parameter space $\Theta$,
PX-EM generates a sequence $\{\theta^{(t)}=R(\theta_*^{(t)}, \alpha
^{(t)})\}$
and converges no slower than the corresponding
EM based on $g(\Xobs, \Xmis;\theta)$; see Liu, Rubin and Wu (\citeyear{LiuRubinWu1998}).

For simplicity and stability, Liu, Rubin and Wu (\citeyear{LiuRubinWu1998}) use
$(\theta^{(t)}, \alpha_0)$ instead of $(\theta_*^{(t)}, \alpha^{(t)})$
for the E~step. As a result, PX-EM shares with EM its E step and modifies
its M step by mapping $(\theta_*^{(t+1)}, \alpha^{(t+1)})$
to the original space $\theta^{(t+1)}=R(\theta_*^{(t+1)}, \alpha^{(t+1)}).$
More precisely, the PX-EM algorithm is defined by replacing the E and M
steps of
EM with the following: 

\begin{description}
\item[\textnormal{\textit {PX-E step.}}] Compute
\begin{eqnarray*}
&&Q\bigl(\theta_*, \alpha|\theta^{(t)}, \alpha_0\bigr)\\
&& \quad
= \mathrm{E} \bigl(\ln g_*(\Xobs, \Xmis;\theta_*, \alpha)
|\Xobs, \theta_*=\theta^{(t)},\\
&& \quad\hphantom{= \mathrm{E} \bigl(\ln g_*(\Xobs, \Xmis;\theta_*, \alpha)
|\Xobs,\theta_*}  \alpha=\alpha_0 \bigr)
\end{eqnarray*}
as a function of $(\theta_*, \alpha)\in\Theta\times{\cal A}$.
\item[\textnormal{\textit {PX-M step.}}] Find
\[
\bigl(\theta_*^{(t+1)}, \alpha^{(t+1)}\bigr)
= \arg\max_{\theta_*, \alpha}Q\bigl(\theta_*, \alpha|\theta^{(t)},
\alpha_0\bigr)
\]
and update
\[
\theta^{(t+1)} = R\bigl(\theta_*^{(t+1)}, \alpha^{(t+1)}\bigr).
\]
\end{description}

Since it is the ordinary EM applied to the parameter expanded
complete-data model,
PX-EM shares with EM its simplicity and stability.
Liu, Rubin and Wu (\citeyear{LiuRubinWu1998}) established theoretical results to show
that PX-EM can converge no slower than EM. Section
\ref{sec:px-robit} uses the robit regression example
to give the statistical interpretation
of Liu, Rubin and Wu (\citeyear{LiuRubinWu1998}) in terms of covariance adjustment.
With the toy example, Section
\ref{sec:ae-px-em} demonstrates that PX-EM can be dramatically
faster than its parent EM.
A discussion
of why PX-EM can perform better than EM is
given in Section \ref{sec:unfolding}.

\subsection{Efficient Analysis of Imputed Missing Data: Robit Regression}
\label{sec:px-robit}
The E step of EM imputes the sufficient statistics $S_{\tau xx'}$ and
$S_{\tau xz}$ with their expectations based on the predictive distribution
of the missing $(\tau_i, z_i)$ data conditioned on the observed data
$\Xobs$ and $\beta^{(t)}$, the current estimate of $\beta$ at the $t$th
iteration. Had the ML estimate of $\beta$, $\hat\beta$, been used to
specify the predictive distribution, EM would have converged 
on the following M step, which in this case performs correct
ML inference. We call the predictive distribution using $\hat\beta$
the correct imputation model.
Before convergence, \textit{that is}, $\beta^{(t)}\neq\hat\beta$,
the E step imputes the sufficient statistics $S_{\tau xx'}$ and
$S_{\tau xz}$ 
using an incorrect imputation model.
The M step also uses a wrong model since it does not
take into account that the data were incorrectly imputed based on
an assumed parameter value $\beta^{(t)}\neq\hat\beta$.
The M step moves the estimate $\beta^{(t+1)}$ toward
$\hat\beta$, but the difference between $\beta^{(t+1)}$ and $\hat
\beta$
can be regarded as bias due to the use of the $\beta^{(t)}$.

The bias induced by the E step can be reduced by making use of
recognizable discrepancies between imputed statistics and their values
under the correct imputation model.
To capture such discrepancies, Liu, Rubin and Wu (\citeyear{LiuRubinWu1998}) considered
parameters that are statistically identified in the complete-data
model but
not in the observed-data model.
These parameters are fixed at their
default values to render the observed-data model identifiable.
In the context of EM for robit regression, these parameters
are the scale parameters of $\tau_i$ and $z_i$,
denoted by $\alpha$ and $\sigma$.
In the observed-data model, they take the
default values $\alpha_0=1$ and $\sigma_0 = 1$.

When activated, the extra parameters are estimated by the M step
and these estimates converge to the default values
to produce ML parameter estimates for the observed
data model.
Thus, in the robit regression model, we identify the default values of
the extra parameters as MLEs:
$\alpha_0=\hat\alpha=1$ and $\sigma_0 = \hat\sigma=1$.
Denote the corresponding
EM estimates by $\alpha^{(t+1)}$ and $\sigma^{(t+1)}$.
The discrepancies between $(\alpha^{(t+1)}, \sigma^{(t+1)})$
and $(\hat\alpha, \hat\sigma)$ reveal the existence of bias
induced by the wrong imputation model.
These discrepancies can be used to adjust the estimate of
the parameter of interest, $\beta$, at each iteration.
This is exactly what PX-EM is formulated to do, and
the resulting algorithm converges faster than the original EM.

Formally, the extra parameter $(\alpha, \sigma)$ introduced to
capture the
bias in the imputed values of $\tau_i$ and $z_i$ is called the
\textit{expansion parameter}. The complete-data model is thus both
data-augmented
as well as parameter-augmented. For correct inference at convergence,
data augmentation is required to preserve the observed-data model
after integrating out missing data.
Likewise, parameter expansion needs to satisfy the observed-data
model preservation condition
(\ref{eq:px-preservation}). In the robit regression model,
let $(\beta_*, \alpha, \sigma)$ be the expanded parameter
with $\beta_*$ playing the same role as $\beta$ in the original
model.
The preservation condition states that for every
expanded parameter value $(\beta_*, \alpha, \sigma)$,
there exists a value of $\beta$ such that
the sampling model of the $y_i$'s obtained from the parameter
expanded model is the same as the original sampling model given $\beta$.
This condition defines a mapping $\beta= R(\beta_*, \alpha, \sigma)$,
the reduction function.
This reduction function is used in PX-EM to adjust the value of $\beta^{(t+1)}$
produced by the M step.

The detailed implementation of PX-EM for robit regression is as follows.
The parameter-expanded complete-data model is obtained by
replacing (\ref{eq:tau}) and (\ref{eq:z}) with
%
\begin{equation}\label{eq:x-tau}
(\tau_i/\alpha)|(\beta_*, \alpha, \sigma) \sim
\operatorname{Gamma}(\nu/2, \nu/2)
\end{equation}
and
%
\begin{equation}\label{eq:x-z}
z_i|(\tau_i,\beta_*, \alpha, \sigma) \sim
\mathrm{N}(x_i'\beta_*, \sigma^2/\tau_i)
\end{equation}
for $i=1,\ldots ,n$. Routine algebraic operation yields the reduction function
%
\begin{eqnarray}\label{eq:reduction}
 \qquad \beta&=& R(\beta_*, \alpha, \sigma)\nonumber
\\[-8pt]
\\[-8pt]  \qquad &=& (\alpha^{1/2}/\sigma)\beta_* \quad
 (\beta_* \in{\cal R}^p; \alpha>0; \sigma>0).
\nonumber
\end{eqnarray}

The expanded parameterization in \eqref{eq:x-tau} and \eqref{eq:x-z}
is a natural choice
if the missing data are viewed as real
and a parameterization is sought that provides a model that is flexible
while preserving the observed data model and allowing the original
parameterization
to be recovered through the reduction function.
For example, if $\tau_i$ is treated as fixed, the model for $z_i$
is a regression model with fixed variance.
Adding $\sigma^2$ and $\alpha$ allows the variance
of $z_i$ and the scale of $\tau_i$ to be estimated freely in the
expanded model.

The sufficient statistics for the expanded parameter
$(\beta_*, \alpha, \sigma)$ now become
%
\begin{eqnarray}\label{eq:x-ss}
S_\tau&=& \sum_{i=1}^n\tau_i, \qquad
S_{\tau xx'} = \sum_{i=1}^n\tau_ix_ix_i',\nonumber
\\[-8pt]
\\[-8pt]
S_{\tau z^2} &=& \sum_{i=1}^n\tau_iz_i^2, \qquad
S_{\tau xz} = \sum_{i=1}^n\tau_ix_iz_i'.
\nonumber
\end{eqnarray}
The complete-data ML estimate of $\beta_*$ is the same as that of
$\beta$ in the original complete-data model.
The complete-data ML estimates of $\alpha$ and $\sigma$ are given by
%
\begin{eqnarray}\label{eq:ml-alpha}
\hat{\alpha}_{{\mathrm{com}}} &=&
\frac{1}{n}S_\tau
  \quad \mbox{and}\nonumber
  \\[-8pt]
  \\[-8pt]
\hat{\sigma}^2_{{\mathrm{com}}} &=&
\frac{1}{n} (
S_{\tau z^2} - S_{\tau xz}S_{\tau xx'}^{-1}S_{\tau xz} ).
\nonumber
\end{eqnarray}

The PX-EM algorithm is simply an EM applied to the parameter expanded
complete-data model with an M step followed by (or modified to contain)
a reduction step. The reduction step
uses the reduction function to map the estimate in the expanded parameter
space to the original parameter space. For the robit example, PX-EM is
obtained by modifying the E and M steps as follows.
\begin{description}
\item[\textnormal{\textit {PX-E step.}}] This is the same as the E step of EM except for
the evaluation of two additional expected sufficient statistics:
\[
\hat{S}_\tau= \mathrm{E} \bigl(S_\tau|\beta=\beta^{(t)},\Xobs
 \bigr)
= \sum_{i=1}^n \hat\tau_i
\]
%
and
\begin{eqnarray*}
\hat{S}_{\tau z^2} &=& \mathrm{E} \bigl(S_{\tau z^2}|\beta=\beta
^{(t)},\Xobs
 \bigr)\\
  &= &n(\nu+1)\\
  &&{} -\nu\sum_{i=1}^n \hat\tau_i
+\sum_{i=1}^n \hat\tau_i x_i'\beta^{(t)}\bigl(2\hat{z}_i-x_i'\beta^{(t)}\bigr),
\end{eqnarray*}
%
where $\hat{\tau}_i$'s and $\hat{z}_i$'s are available from the
E step of EM.
\item[\textnormal{\textit {PX-M step.}}] Compute
$\hat\beta_*=\hat{S}_{\tau xx'}^{-1}\hat{S}_{\tau xz}$,
$\hat\sigma_*^2= n^{-1}\times\break  (
\hat S_{\tau z^2}
- \hat S_{\tau xz}\hat S_{\tau xx'}^{-1}\hat S_{\tau xz} ),$
and $\hat\alpha_*= n^{-1}\hat{S}_{\tau}$ and then use
the reduction to obtain
$\hat\beta^{(t+1)} =
(\hat\alpha_*^{1/2}/\hat\sigma_*) \hat\beta_*$.
\end{description}

For a numerical example, consider the data of Finney (\citeyear{Finney1947}), which consist
of 39 binary responses denoting the presence ($y=1$) or absence
($y=0$) of vaso-constriction of the skin of the subjects after
inspiration of a volume $V$ of air at inspiration rate $R$.
The data were obtained from repeated measurements on three
individual subjects, the numbers of observations per subject
being 9, 8 and 22. Finney (\citeyear{Finney1947}) found no evidence of inter-subject
variability, treated the data as 39 independent observations, and analyzed
the data using the probit regression model with $V$ and $R$ in the logarithm
scale as covariates. This data set was also analyzed by
Liu (\citeyear{Liu2004}) to illustrate robit regression.
Due to three outlying observations, the MLE of the degrees of freedom
$\nu$ is very small, $\hat\nu=0.11$.

\begin{figure*}

\includegraphics{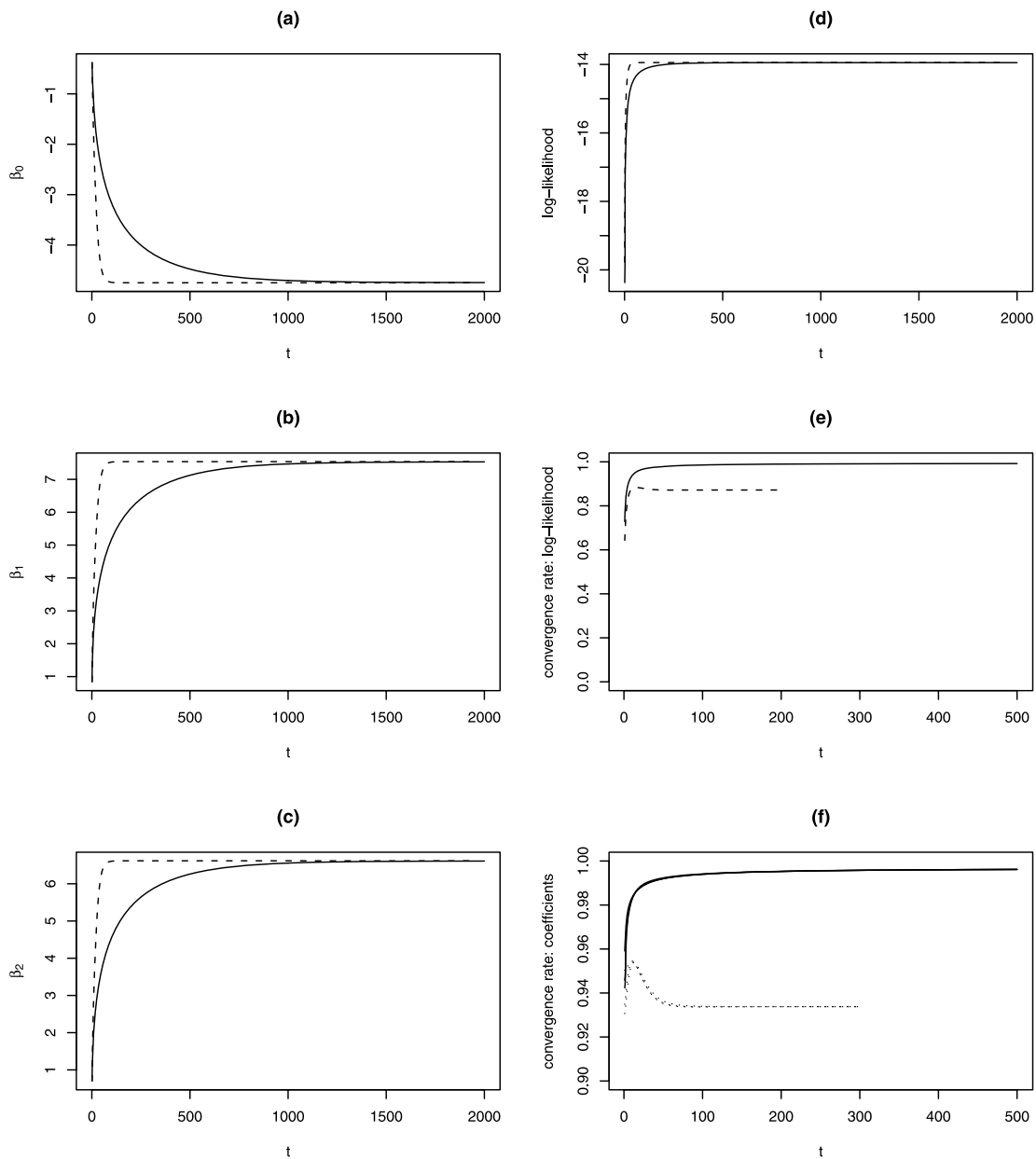}

\caption{EM (solid) and PX-EM (dashed) sequences of
the regression coefficients $\beta_0$ \textup{(a)}, $\beta_1$ \textup{(b)},
$\beta_2$ \textup{(c)}, and log-likelihood in the robit regression with $x=(1,
\ln(V), \ln(R))$.
The rates of convergence of EM (solid) and PX-EM (dashed)
are shown in \textup{(e)} by
$|\ell^{(t+1)}-\ell^{(\infty)}|/|\ell^{(t)}-\ell^{(\infty)}|$,
where $\ell^{(t)}$ denotes the log-likelihood value at the $t$th iteration,
and in \textup{(f)} by
$|\beta_j^{(t+1)}-\beta_j^{(\infty)}|/|\beta_j^{(t)}-\beta
_j^{(\infty)}|$ for
$j=0, 1$ and $2$.}
\label{fig:robit}
\end{figure*}

Here we use this data set
with $\ln(V)$ and $\ln(R)$ as the covariates and take
the fixed $\nu=2$ as a numerical example to compare EM and PX-EM.
Numerical results comparing the rates of convergence of EM and
PX-EM are displayed in Figure \ref{fig:robit}.
PX-EM shows a clear and dramatic convergence gain over EM.
For convenience we choose to report the detailed results
over iterations. The algorithms were coded in R, which
makes CPU comparison unreliable.
Since extra computation for the PX-EM implementation
is minor, we believe the same conclusion holds in terms of CPU times.

\subsection{PX-EM with Fast Convergence: The~Toy~Example}
\label{sec:ae-px-em}

The model $X|(Z,\lambda)\sim\operatorname{Binomial}(Z, \pi)$ may not
fit the imputed value of missing data $Z$ very well
in the sense that $X/\hat{Z}$ is quite different from $\pi$.
This mismatch can be used to adjust $\lambda^{(t+1)}$.
To adjust $\lambda^{(t+1)}$, we activate $\pi$ and let $\alpha$
denote the activated parameter with $\alpha_0=\pi$.
Now the parameter-expanded complete-data model becomes
\[
Z|(\lambda_*, \alpha) \sim\operatorname{Poisson}(\lambda_*)
\]
and
\[
X|(Z, \lambda_*, \alpha) \sim\operatorname{Binomial}(Z, \alpha),
\]
where $\lambda_* >0$ and $\alpha\in(0,1)$.
If the missing data were treated as being observed,
this model allows the mean parameters for both $X$ and $Z$ to be
estimated.
The two corresponding observed-data models are
$\operatorname{Poisson}(\lambda\pi)$
and $\operatorname{Poisson}(\lambda_*\alpha)$, giving the reduction
function
%
\begin{equation}\label{eq:ae-reduction}
\lambda= R(\lambda_*,\alpha) = \frac{\alpha}{\pi}\lambda_*.
\end{equation}

The complete-data sufficient statistics are $Z$ and $X$.
The complete-data ML estimates of $\lambda_*$ and $\alpha$
are given by
%
\begin{equation}\label{eq:lh-com}
\hat\lambda_{*,\mathrm{com}} = Z
 \quad \mbox{and} \quad
\hat\alpha_{\mathrm{com}} = \frac{X}{Z}.
\end{equation}
The resulting PX-EM has the following E and M steps:
\begin{description}
\item[\textnormal{\textit {PX-E step.}}] This is the same as the E step of EM.
\item[\textnormal{\textit {PX-M step.}}] Replace $Z$ in (\ref{eq:lh-com})
with $\hat{Z}$ to obtain $\lambda_*^{(t+1)} = \hat{Z}$ and
$\alpha^{(t+1)} = X/\hat Z$. Update $\lambda$ using the
reduction function and obtain
\[
\lambda^{(t+1)} = \frac{ X}{\pi\hat Z} \hat{Z} = \frac{X}{\pi}.
\]
\end{description}
The PX-EM algorithm in this case converges in one step.
Although artificial, this toy example shows again that
PX-EM can converge dramatically faster than its parent EM.

\section{Unfolding the Mystery of PX-EM}
\label{sec:unfolding}

The statistical interpretation in terms of covariance adjustment,
explained by the robit example above and the Student-$t$ example in
Liu, Rubin and Wu (\citeyear{LiuRubinWu1998}), and the theoretical results of Liu, Rubin and Wu
(\citeyear{LiuRubinWu1998})
help reveal the PX-EM magic.
To further unfold the mystery of PX-EM, we discuss
the nonidentifiability of expanded parameters in the observed-data model
in Section \ref{sec:how-to}
and take a look at PX-EM from
the point of view of efficient data augmentation
in Section~\ref{sec:eda}.

\subsection{Nonidentifiability of Expanded Parameters and
Applicability of PX-EM}
\label{sec:how-to}

It is often the case in PX-EM that,
even though the expanded parameter $(\theta_*, \alpha)$ is
identifiable from
$Q(\theta_*,\alpha|\theta^{(t)},\alpha_0)$
(the expected parameter-expanded complete-data log-likelihood),
it is not identifiable
from the corresponding observed-data loglikelihood
\[
L_*(\theta_*,\alpha) = \ln f_*(\Xobs; \theta_*,\alpha).
\]
It is helpful to consider $L_*(\theta_*,\alpha)$
for understanding PX-EM, as the PX-M step
directly increases $L_*(\theta_*,\alpha)$ through
maximizing $Q(\theta_*,\alpha|\theta^{(t)},\alpha_0)$.
Naturally, from a mathematical point of view,
unfolding the actual likelihood
in the larger or expanded parameter space
$\Theta\times{\cal A}$
shows how PX-EM steps can lead to increases in the likelihood function faster
than can moves in the original space $\Theta$.

\subsubsection{The observed-data log-likelihood surface over $\Theta
\times{\cal A}$}

The observed-data log-likelihood, as a function of $(\theta_*, \alpha)$,
is determined by the actual log-likeli\-hood $L(\theta)=\ln f(\Xobs
;\theta)$
with $\theta$ replaced by $\theta=R(\theta_*,\alpha)$ so that
%
\begin{equation}\label{eq:L.px} \qquad
L_*(\theta_*,\alpha) = L(R(\theta_*,\alpha)) \quad
 \bigl((\theta_*, \alpha) \in\Theta\times{\cal A}\bigr).
\end{equation}
Thus, each point $\theta\in\Theta$ corresponds to a subspace
$\{(\theta_*, \alpha) \in\Theta\times{\cal A}, R(\theta_*,\alpha
) = \theta\}$,
over which $L_*(\theta_*,\alpha)$ is constant.

\begin{figure}

\includegraphics{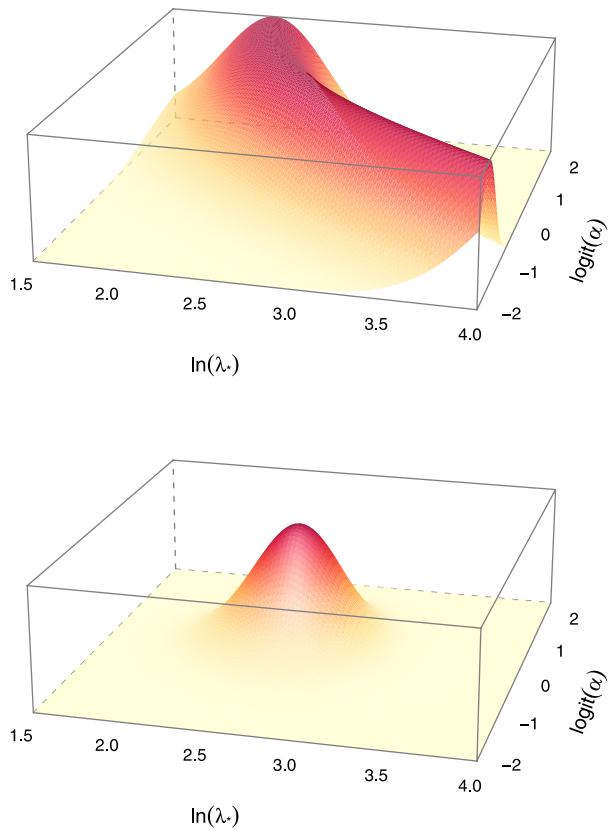}

\caption{Perspective plots of
the parameter-expanded observed-data log-likelihood
$L(\lambda_*,\alpha)$ \textup{(top)} and
the parameter-expanded complete-data log-likelihood
$Q(\lambda_*,\alpha|\lambda^{(t)})$ \textup{(bottom)} in the toy example with
$X=8$, $\pi= 0.25$, and $\lambda^{(t)}=8$.} \label{fig:toypersp}
\end{figure}

\begin{figure}

\includegraphics{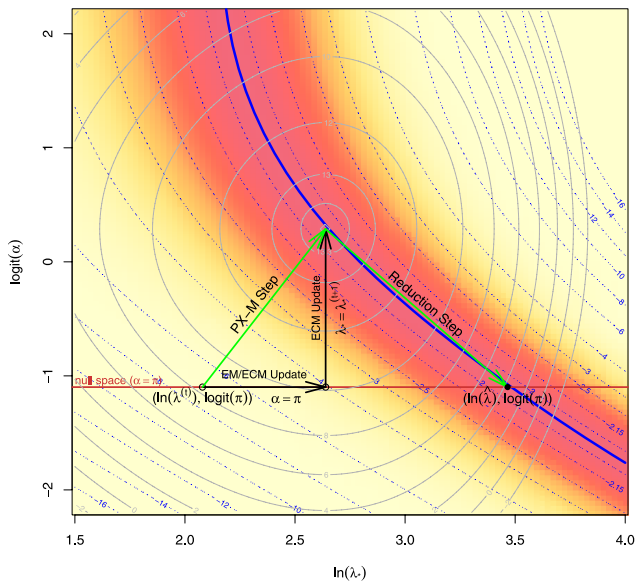}

\caption{PX-EM for the toy example with $X=8$, $\pi= 0.25$, and $\lambda^{(t)}=8$.
The parameter-expanded observed-data log-likelihood function
$L(\lambda_*, \alpha)$
is shown by shading and dashed contours
with a maximum along the ridge indicated by
a solid thick line.
The expected parameter-expanded complete-data log-likelihood $Q(\lambda
_*, \alpha|
\lambda^{(t)})$ is shown by the ellipse-like solid contours.
In this example, maximization of $Q(\lambda_*, \alpha| \lambda^{(t)})$
over $(\lambda_*,\alpha)$ can be obtained in two conditional
maximization steps, labeled as the two ECM updates.
The PX-M step moves to a point on the ridge of $L(\lambda_*, \alpha)$,
and the subsequent reduction-step moves this point along
the the ridge of $L(\lambda_*, \alpha)$ to the point with $\alpha
=\pi$.} \label{fig:toy}
\end{figure}

For example, when $\theta$ and $\alpha$ are one-dimensional parameters,
$L(\theta)$ can be represented by a curve in the two-dimensional space
$\Theta\times L(\Theta)$, whereas $L_*(\theta_*,\break\alpha)$ is a family
of curves indexed by $\alpha$.
The family of curves $L_*(\theta_*, \alpha)$ form a surface
in the style of a mountain range in the three-dimensional space $\Theta
\times{\cal A}\times L(\Theta)$.
For the toy example,
this is depicted in Figure~\ref{fig:toypersp}
by the top panel 3-D perspective plot
and in Figure~\ref{fig:toy}
by the image with dashed contours or ``elevation'' lines.
The mode of $L(\theta)$ now becomes a set of modes of the same
``altitude,'' one for each fixed $\alpha$. That is,
the mode of $L(\theta)$ is expanded into the ``ridge'' shown, for
example, by the
thick line in Figure \ref{fig:toy}.

\subsubsection{Likelihood maximization in PX-EM}
The E step in PX-EM implicitly computes a family of
expected complete-data log-likelihood functions, which are the
$Q$-functions used in \eqref{eq:Q-fun},
over the original parameter space indexed by the expansion
parameter $\alpha$.
This is because PX-EM introduces no additional or different missing data
in the larger complete-data model.
In other words, the parent E step effectively computes
a surface over $\Theta\times{\cal A}$ that can be used
as a $Q$-function to approximate the expanded loglikelihood
$L_*(\theta_*,\alpha)$ defined in (\ref{eq:L.px}).
This $Q$-function for the toy example is shown
in Figure \ref{fig:toypersp}
by the bottom panel 3-D perspective plot
and in Figure \ref{fig:toy} by the nearly-elliptical contours.
For this one-step convergence PX-EM example,
the mode of this $Q$-function is on the ridge of the expanded loglikelihood
$L_*(\theta_*,\alpha)$.
We note that this is typically not the case in more realistic examples.
In the general case, the mode of the $Q$-function
would typically be located on one elevation line that is
higher than the elevation line where the update $(\theta^{(t)}, \alpha
_0)$ found
by EM is located.

Somewhat surprisingly, any such $Q$-function for each
fixed $\alpha$ is EM-{valid}.
By \textit{EM-valid}, we mean that increasing the $Q$-function results in
an increase of the actual likelihood in the expanded space
and thereby in the original space after the reduction step.
This is due to two facts: (i) the joint $Q$-function
is EM-valid for $L_*(\theta_*,\alpha)$ and, thus, for $L(\theta)$ as well,
and (ii) an M~step with any fixed $\alpha$, which finds
\[
\theta^{(t+1)}_* = \arg\max_{\theta_*}
Q\bigl(\theta_*, \alpha|\theta^{(t)},\alpha_0\bigr),
\]
followed by the reduction $\theta^{(t+1)}=R(\theta^{(t+1)}_*,\alpha)$
is simply a conditional maximization step.
Additionally, in the context of the ECM algorithm of Meng and Rubin (\citeyear{MengRubin1993}),
the parent EM is an incomplete ECM with only one single CM step over
$\Theta\times{\cal A}$.
This relationship is explored in greater detail in the next section.

\subsubsection{PX-EM vs. (efficient) ECM over $\Theta\times{\cal A}$}
In theory,
PX-EM has a single M step over the entire space $\Theta\times{\cal A}$.
Note that
\[
\max_{(\theta_*,\alpha)}
Q\bigl(\theta_*, \alpha|\theta^{(t)},\alpha_0\bigr)
=
\max_\alpha\max_{\theta_*}
Q\bigl(\theta_*, \alpha|\theta^{(t)},\alpha_0\bigr).
\]
When
\[
\hat\theta_*^{(t+1)} = \arg\max_{\theta_*}
Q\bigl(\theta_*, \alpha|\theta^{(t)},\alpha_0\bigr)
\]
does not depends on $\alpha$, as is often the case in
many PX-EM examples, the PX-M step is equivalent
to a cycle of two CM steps: one is the M step of EM,
and the other updates $\alpha$ with $\theta_*$ fixed at
$\theta_*^{(t+1)}$.
This version of ECM for the toy example
is illustrated in Figure \ref{fig:toy}.
In this case, ECM is efficient for it generates the PX-EM update.

To summarize, denote by $\mathrm{ECM}_{\{\alpha,\theta_*\}}$ the above~version of ECM
over $\Theta\times{\cal A}$.
Typically,
the algorithms can then be ordered in terms of performance as
%
\begin{equation}
\mathrm{EM}  \preceq  \mathrm{ECM}_{\{\alpha,\theta_*\}}  \preceq
  \mathrm{PX-EM}
\end{equation}
over $\Theta\times{\cal A}$.
It should be noted that by \textit{typically}, we mean the conclusion
is reached 
in an analogy with
comparing the EM algorithm and the Generalized EM algorithm
(GEM) (Dempster, Laird and Rubin, \citeyear{DLR}), that is,
EM typically converges faster than GEM, but counter
examples exist; see, \textit{for example}, Section 5.4 
of van Dyk and Meng (\citeyear{vanDykMeng2010}) and the alternative explanation
from an ECME point of view
in Section 4.3 of Liu and Rubin (\citeyear{LiuRubin1998})
on why ECM can be faster than EM.
To elaborate it further with our robit example,
it may be also interesting to note that
when the reduction function (\ref{eq:reduction}) is modified
by replacing the adjustment factor $(\alpha^{1/2}/\sigma)$
with $(\alpha/\sigma)$,
a typo made in the earlier
versions of the PX-EM for the robit regression model, the resulting
(wrong) PX-EM converges actually faster than the (correct) PX-EM
for the numerical example in Section 3.2.
In general, more efficiency can~be gained by
replacing the CM step of ECM over $\alpha$ with a CM step
maximizing the corres\-ponding actual constrained likelihood in
the parameter expanded space.
This is effectively a parameter-expanded ECME algorithm;
see such an example for the Student-$t$ distribution given in Liu (\citeyear{Liu1997}).
More discussion on ECME and other
state-of-the-art methods for accelerating the EM algorithm
can be found in He and Liu (\citeyear{he2009dynamic}). Their discussion on the method
termed SOR provides a relevant explanation why the
above wrong PX-EM and other wrong PX-EM versions, such as the one
using the wrong reduction function
$
\beta = (\alpha/\sigma^2)\beta_*
$
in the numerical robit example, can converge
faster than the correct PX-EM.

Perhaps most importantly, the above discussion
further explains why PX-EM can perform better than EM can, and
unfolds the mystery of PX-EM, in addition to
the covariance adjustment interpretation.
\subsection{Efficient Data Augmentation via Parameter Expansion}
\label{sec:eda}


Meng and van Dyk (\citeyear{MengVanDyk1997}) consider efficient data augmentation for
creating fast converging algorithms.
They search for efficient augmenting schemes by working with the
fraction of missing-data information. Here we show that PX-EM
can also be viewed as an alternative way of doing efficient
data augmentation. Unlike Meng and van Dyk (\citeyear{MengVanDyk1997}), who find a fixed
augmenting scheme
that works for all EM iterations,
the following procedure is a way to
choose an adaptive augmenting scheme for each EM iteration.
Rather than control the fraction of missing-data information,
this procedure reduces bias through the expansion parameter.
For the sake of clarity, we use the artificial example of Section \ref{sec:ae}
to make our argument.

Consider the parameter-expanded complete-data likelihood
obtained from (\ref{eq:ae-likelihood}) by activating
$\alpha_0=\pi$, \textit{that is},
\begin{eqnarray}
\frac{\lambda_*^{Z}}{Z!}e^{-\lambda_*}
\frac{Z!}{X!(Z-X)!} \alpha^{X}(1-\alpha)^{Z-X}\nonumber\\
 \eqntext{(\lambda_*>0; 0<\alpha<1),}
\end{eqnarray}
which has the canonical representation
\begin{eqnarray}
h(X,Z) c(\lambda_*, \alpha) e^{Z\ln[\lambda_*(1-\alpha)]
+ X\ln \alpha/(1-\alpha) }\nonumber\\
 \eqntext{(\lambda_*>0; 0<\alpha<1).}
\end{eqnarray}
Thus, when fixed at the given value, $\pi$, for identifiability,
the complete-data ML estimate $\hat\alpha=X/Z$ plays the role
of an ancillary statistic; see Ghosh, Reid and Fraser (\citeyear{GhoshReidFraser2010}) 
for an introduction to ancillary statistics.
With the correct imputation model, or at convergence, the imputed value
$\hat{Z}$ satisfies
%
\begin{equation}\label{eq:constrained-ancillary}
\pi= \frac{X}{\hat{Z}}  \quad \mbox{or} \quad
\hat{Z} = \frac{X}{\pi}.
\end{equation}
Thus, we can consider modifying the E step of EM
to produce an imputed statistic $\hat{Z}$ that satisfies
(\ref{eq:constrained-ancillary}).

In the context of PX-EM, the current estimate
$\lambda^{(t)}$ corresponds to the following subset of the expanded
parameter space:
%
\begin{eqnarray}\label{eq:MLsubset}
\Omega_*^{(t)}&\equiv&
\bigl\{(\lambda_*, \alpha)\dvtx
R(\lambda_*, \alpha) = R\bigl(\lambda^{(t)},
\alpha_0\bigr)\bigr\}\nonumber
\\[-8pt]
\\[-8pt]
&=&\bigl\{(\lambda_*, \alpha)\dvtx  \lambda^{(t)}\pi= \alpha\lambda_*\bigr\}.
\nonumber
\end{eqnarray}
Thus, we can use the imputation
model defined by the parameter-expanded complete-data model
conditioned on an arbitrary point
$(\tilde\lambda_*, \tilde\alpha) \in\Omega_*^{(t)}$.
For efficient data augmentation, we choose a particular
point $(\tilde\lambda_*, \tilde\alpha) \in\Omega_*^{(t)}$, if it exists,
so that (\ref{eq:constrained-ancillary}) holds.
Since
\[
\hat{Z} = \mbox{E} (Z|X, \lambda_*, \alpha )
= X + \lambda_*(1-\alpha),
\]
to obtain the desired imputation model, we solve
\begin{eqnarray*}
X + \tilde\lambda_*(1-\tilde\alpha) & = & \frac{X}{\pi},\\
\lambda^{(t)}\pi& = & \tilde\alpha\tilde\lambda_*
\end{eqnarray*}
for $(\tilde\lambda_*, \tilde\alpha)$.
This system of equations has the solution
\[
\tilde\lambda_* = X \frac{1-\pi}{\pi} + \lambda^{(t)} \pi
 \]
  and
  \[
\tilde\alpha=\frac{\lambda^{(t)} \pi}{X((1-\pi)/\pi
)+\lambda^{(t)} \pi}.
\]
The E step of the EM algorithm based on the corresponding imputation model
produces $\hat{Z} = X/\pi$. The following M step of EM gives
$\lambda^{(t+1)} = \hat{Z} = X/\pi$.

The resulting EM algorithm is effectively the PX-EM algorithm.
This implies that PX-EM can be understood from the perspective of
efficient data augmentation via parameter expansion.
Similar arguments can be made for other PX-EM examples having
imputed ancillary statistics.
In the general case, such an efficient data augmentation
amounts to modifying imputed complete-data sufficient statistics and
can be viewed as re-imputation of missing
sufficient statistics.

\section{Discussion}
\label{sec:discussion}

Gelman (\citeyear{Gelman2004}) notes that ``progress in statistical computation
often leads to advances in statistical modeling,''
which opens our eyes to the broader picture.
Statistical interpretations of EM and PX-EM reveal that
statistical thinking can aid in understanding and
developing iterative algorithms.
It seems natural to apply
fundamental concepts from statistical inference
to address statistical problems such as
ML estimation and Bayesian estimation
(see, e.g., Liu and Wu, \citeyear{LiuWu1999}; van Dyk and Meng, \citeyear{vanDykMeng2001};
Qi and Jaakkola, \citeyear{QiJaakkola2007}; Hobert and Marchev, \citeyear{HobertMarchev2008}).
A recent example is the work of  Yu and Meng (\citeyear{YuMeng2008,YuMeng2010}),
which uses relationships motivated by the concepts of ancillarity and
sufficiency in order to find optimal parameterizations for
data augmentation algorithms used in Bayesian inference.
However, statistical thinking can also be helpful
for general-purpose optimization algorithms
such as in the improvements to the quasi-Newton algorithm developed by
Liu and Vander Wiel (\citeyear{LiuVanderWiel2007}).

Thinking outside the box, 
here we briefly discuss other potential
applications of parameter expansion to statistical inference.
The fundamental idea of PX-EM---the use of expanded parameters to capture
information in data---leads immediately to a possible
application of parameter expansion for ``dimension-matching''
in Fisher's conditional inference and
fiducial inference 
(see, e.g.,  Fisher, \citeyear{Fisher1973}), where difficulties arise when the
dimensionality of the minimal
\mbox{sufficient} statistics is larger than the number of free parameters to be
inferred.
It is well known that, while attempting to build a solid foundation for
statistical inference, the ideas behind Fisher's fiducial inference
have not been well developed.
Nevertheless, it is expected that parameter expansion can be useful in
developing new ideas for statistical inference.
For example, a Dempster--Shafer or fiducial-like method, called the
inferential model (IM) framework,
has been proposed by Zhang and Liu (\citeyear{ZhangLiu2010}) and Martin, Zhang and Liu
(\citeyear{MartinZhangLiu2010}).
Of particular interest is the parameter expansion technique proposed by
Martin, Hwang and Liu (\citeyear{MartinHwangLiu2010}) for what they call weak marginal inference
using IMs.
Using this parameter expansion technique, they provide
satisfactory resolutions to the famous Stein's paradox
and the Behrens--Fisher problem.

Although brief, the above discussion shows that parameter expansion has
the potential to contribute to a variety of applications in computation
and statistical inference.
To conclude this review article, we speculate on
one possible application of parameter expansion to the method
of maximum likelihood for which the EM algorithm has proven
to be a useful computational tool.
The prospect of applying general statistical ideas to computational problems
has also led us to thinking about model checking or goodness of fit
to solve the unbounded likelihood problem in fitting Student-$t$ and
mixture models, for which EM is often the first choice.
In the case with unbounded likelihood functions, for example,
a high-likelihood model may not fit the
observed data well and then
inferential results can be nonsensical.
It would be interesting to see if the general idea of
parameter expansion for efficient inference can be extended
for ``valid inference'' as well.
However, it is not our intention here to discuss these
open problems in depth.
Based on the past success in this area,
it can be expected that parameter expansion methods will continue to
aid computation and inference. 

\section*{Acknowledgments}
The authors thank the editors and their reviewers for their helpful
comments and suggestions
on earlier versions of this article.
Chuanhai Liu was partially supported by NSF
Grant DMS-10-07678.

\end{document}